\documentclass[12pt,a4paper]{article}
\usepackage[utf8]{inputenc}
\usepackage[T1]{fontenc}
\usepackage{amsmath,amssymb,bbm}
\usepackage{graphicx}

\begin{document}
\textwidth=135mm
 \textheight=200mm
\begin{center}
{\bfseries The same key to different doors - temperature puzzles
\footnote{{\small Lecture at the 32th Max-Born-Symposium and HECOLS workshop on "Three Days of Phase Transitions in Compact Stars, Heavy-Ion Collisions and Supernovae", Institute of Theoretical Physics, University of Wroc\l{}aw, Wroc\l{}aw, Poland, February 17--19, 2014.}}}
\vskip 5mm
L.~Turko$^{\dag}$
\vskip 5mm
{\small {\it $\dag$ Uniwersytet Wroc\l{}awski, Instytut Fizyki Teoretycznej,
Wroc\l{}aw, Poland}}\\
\end{center}
\vskip 5mm
\centerline{\bf Abstract}
The notion of temperature in many body elementary particle processes is in a common use for decades. Thermal models have become simple and universal effective tools to describe particle production -- not only in high energy  heavy ion collisions but also in high energy elementary particle collisions. We perform a critical analysis of the  temperature concepts  in such processes.  Although the temperature concept is a very useful tool, nevertheless it should be used with the care, taking into account that usually it is just model dependent fitted parameter.
\vskip 10mm
\section{\label{sec:intro}Introduction}
There is the famous article \cite{wigner} by E.P. Wigner  based on the lecture he delivered at New York University in May 1959. You can read there about \emph{"that uncanny usefulness of mathematical concepts that raises the question of the uniqueness of our physical theories"} and that \emph{"We are in a position similar to that of a man who was provided with a bunch of keys and who, having to open several doors in succession, always hit on the right key on the first or second trial. He became skeptical concerning the uniqueness of the coordination between keys
and door."}

That was about mathematics. But some suspicions concerning universal keys can be raised not only to mathematical concepts. We are in possession of some universal key whose usefulness becomes more and more universal up to the point when another question arises -- is this a real key?

\section{\label{sec:fitting}Fitting the keys}
Discovery of pions in 1948, new light strongly interacting particles, copiously produced and observed in cosmic ray processes, created a new situation in the theoretical description of production processes. There was a need to deal with the global description of these processes, leaving temporally aside microscopic subtleties of interactions. The natural tools available at that time was to use some analogy to concepts used in  many body classical physics. There were attempts to treat mesonic cloud as a hydrodynamical medium \cite{Heisenberg  : 1949ah} or as a kind of atmospheric pion gas or fluid surrounding nucleons.  This pion fluid would be set in a kind of turbulent motion in the course of a high energy collision of two nucleons. This turbulence would govern the distribution of energy among different excited states

The concept of temperature was being introduced in elementary particle physics, in a more or less systematic way, many times. It seems, however, that the first, who applied there this notion to particle physics was H.~Koppe \cite{Koppe:1948az, Koppe:1949zz}. He treated a nucleus  as a "black body" with regard to mesonic radiation. This made possible to calculate the probability for emission of a meson by statistical methods.

Then, two years later, has appeared famous Fermi Model \cite{Fermi:1950jd} where temperature was introduced in a more systematic way. The concept of the statistical equilibrium was then used to describe a high energy collision. Fermi wrote there:

\emph{When  two  nucleons  collide  with  very  great  energy  in  their  center  of  mass
system  this  energy  will  be  suddenly  released  in  a  small  volume  surrounding  the
two  nucleons.  (...)  all  the  portion  of space  occupied  by  the  nucleons  and  by  their  surrounding
pion  field  will  be  suddenly  loaded  with  a  very  great  amount  of  energy.  Since the
interactions  of the  pion  field  are strong we  may expect that rapidly this energy
will  be  distributed  among  the  various  degrees  of freedom  present  in  this  volume
according  to  statistical  laws.  One  can  then  compute  statistically  the  probability
that  in  this  tiny  volume  a  certain  number  of pions  will  be  created  with  a  given
energy  distribution.  It is  then  assumed  that  the  concentration  of  energy  will
rapidly  dissolve  and  that the  particles  into  which  the  energy  has  been  converted
will  fly  out  in  all  directions.(...) First  of  all  there  are- conservation  laws  of  charge  and  of
momentum  that  evidently  must  be  fulfilled
}

The main idea of statistical model as sketched in by Fermi \cite{Fermi:1950jd} remains still valid. Let's consider the probability $\mathcal{P}_n(i\to f)$ to produce $n-$particle state
\begin{equation}\label{prob-n}
{\mathcal{P}_n(i\to f)=
  \int\prod\limits_{f=1}^n\frac{d^3p_f}{(2\pi)^3}\frac{1}{2E_f}|\langle p'_1,\dots
p'_n|\mathcal{S}|i\rangle|^2 \delta(P_i-\sum\limits_f p_f)}
\end{equation}
One can clearly separate here the dynamical part
\[|\langle p'_1,\dots
p'_n|\mathcal{S}|i\rangle|^2\]
and the kinematical part of the process
\[\delta(P_i-\sum\limits_f p_f)\prod\limits_{f=1}^n\frac{d^3p_f}{(2\pi)^3}\frac{1}{2E_f}\]%
With the increasing number of final particles the number of degrees of freedom increases much more. We are not able to measure all of them and there is also no need to do it. Only some global quantities are measured. We are in the situation when
\begin{itemize}
  \item measurable quantities are much less detailed than $\langle p'_1,\dots
p'_n|\mathcal{S}|i\rangle$
\item with the integration over a large region of the phase space the dynamical
details are averaged and only a few parameters remains
\item restricted knowledge of $\langle p'_1,\dots
p'_n|\mathcal{S}|i\rangle$ is not needed
\end{itemize}
Then, there is a place for statistical physics.

Our probability \eqref{prob-n} can be written as
\[P_n=\bar{S}_n\,\mathcal{R}_n\]
with the constant averaged value $\bar{S}_n$ of the $S$ matrix element and with the exact kinematical part
\begin{equation}\label{kinem-n}
\mathcal{R}_n=\int\prod\limits_{f=1}^n\frac{d^3p_f}{(2\pi)^3}\frac{1}{2E_f}\delta(P_i-\sum\limits_f p_f)
\end{equation}

The crucial point is here that these arguments work only if the thermodynamic equilibrium is reached.

Rolf Hagedorn was the first who systematically analyzed high energy phenomena using all tools of statistical
physics \cite{Hagedorn:1965st, Hagedorn:1970gh}. He also introduced the concept of \emph{the limiting temperature} based then on the statistical bootstrap model. This made possible the introduction of the possibility of the phase transition and phase structure of the hadronic matter. 

The spirit and the philosophy of the statistical approach remains the same as in the standard approach but
ingredients of statistical models used in high energy problems are different. The
main difference is that a  number of particles is not longer conserved so we have no
chemical potentials related to that quantity. The only nontrivial chemical
potentials are those related to conserved charges, so the role of internal
symmetries is a crucial one. As was stated in \cite{Hagedorn:1970gh} \emph{"Symmetries,  not  material  particles  are  fundamental".}

For the simplest case of an ideal hadron gas in thermal and chemical
equilibrium, which consists of $l$ species of particles, energy density $\epsilon$,
baryon number density $n_{B}$, strangeness density $n_{S}$ and electric charge
density $n_(Q)$ read ($\hbar=c=1$ always) one gets equations

\begin{subequations}\label{eqstate}
  \begin{align}
\epsilon = \frac{1}{2\pi^{2}} \sum_{i=1}^{l} (2s_{i}+1)
\int\limits_{0}^{\infty}dp\,{ \frac{ p^{2}E_{i} }{ \exp \left\{ {{ E_{i} - \mu_{i}
} \over T} \right\} + g_{i} } }
&\ ,\\
n_{B}= \frac{1}{2\pi^{2}} \sum_{i=1}^{l} (2s_{i}+1) \int\limits_{0}^{\infty}dp\,
\frac{p^{2}B_{i}}{ \exp \left\{ {{ E_{i} - \mu_{i} } \over T} \right\} + g_{i} }
&\ ,\\
n_{S}=\frac{1}{2\pi^{2}} \sum_{i=1}^{l} (2s_{i}+1) \int\limits_{0}^{\infty}dp\, \frac{
p^{2}S_{i} }{ \exp \left\{ {{ E_{i} - \mu_{i} } \over T} \right\} + g_{i} }
&\ ,\\
n_Q=\frac{1}{2\pi^{2}} \sum_{i=1}^{l} (2s_{i}+1) \int\limits_{0}^{\infty}dp\,\frac{
p^{2}Q_{i}}{ \exp \left\{ {{ E_{i} - \mu_{i} } \over T} \right\} + g_{i} }
&\ .
\end{align}
\end{subequations}
where $E_{i}= ( m_{i}^{2} + p^{2} )^{1/2}$ and $m_{i}$, $B_{i}$, $S_{i}$, $\mu_{i}$,
$s_{i}$ and $g_{i}$ are the mass, baryon number, strangeness, chemical potential,
spin and a statistical factor of specie $i$ respectively (we treat an antiparticle
as a different specie).

And $\mu_{i} = B_{i}\mu_{B} + S_{i}\mu_{S} + Q_{i}\mu_{Q}$,
where $\mu_{B}$, $\mu_{S}$, and  $\mu_{Q}$ are overall baryon number and strangeness
chemical potentials respectively.

To get particle yields one should consider also entropy density $n(s)$
\begin{equation}\label{entropy}
s={1 \over {6\pi^{2}T^{2}} } \sum_{i=1}^{l} (2s_{i}+1) \int\limits_{0}^{\infty}dp\,
{ {p^{4}} \over { E_{i} } } { { (E_{i} - \mu_{i}) \exp \left\{ {{ E_{i} - \mu_{i} }
\over T} \right\} } \over { \left( \exp \left\{ {{ E_{i} - \mu_{i} } \over T}
\right\} + g_{i} \right)^{2} } }\ .
\end{equation}

These equations, enriched by unstable particles effects, form a basic for successful
calculations \cite{brs} of relativistic heavy ion production processes concerning
particle yields and rates.
\section{Temperature, which temperature?}
Although temperature appeared a quite successful tool to characterize high energy hadronic collision, there are still discussions related to the physical interpretation of this concept. The temperature, as it was introduced in the classical physics based on the direct contact of the measuring device - thermometer - with the given object. It was also tacitly assumed the the thermometer would be small enough to not change the thermodynamic characteristic of the object and the result would be obtained in the state of the thermal equilibrium between the object and the thermometer. This quantity, measured by the direct contact is called the physical temperature. 

No-thermometer measurements of the quantity \emph{called} "temperature" are based on a given model assumptions. Their relations to the physical temperature depends on the validity of the assumed model and on the very existence of the physical temperature of the system. The situation becomes even more complicated in the case of many particle  quantum systems as e.g. multiproduction processes where you deal with nontrivial (mixed states) density matrix. Because of impossibility of the full microscopic description one uses \emph{relevant} entropy $S_\mathbb{X}(A_1,A_2,...)$ with the entropy maximized with respect to the set $\mathbb{X} = \{A_1,A_2,...\}$ of relevant macro-variables. The relevant entropy is maximized under constrains $\langle \hat A_i\rangle = A_i$ 
    \begin{equation}\label{entropy-rel}
    S_\mathbb{X}(A_1,A_2,...) = -{\underset{\varrho}{\max}}\ \mbox{Tr\,}\varrho\ln\varrho\,;\qquad \mbox{Tr}\,\varrho\hat A_i = A_i\,\,.
    \end{equation}
    The maximum is set over all possible distributions $\varrho$ satisfying constraints $\mbox{Tr}\,\varrho\hat A_i = A_i$.

 This relevant entropy takes into account only information connected with relevant variables. If one of the relevant variables is taken energy then temperature is defined as
 \[\frac{1}{T} = \frac{\partial  S_\mathbb{X}}{\partial E}\]

This is in fact just the temperature widely used in thermal hadronic models. It is obvious that the entropy is unique for the given set of relevant variables. For different choice of relevant variables the temperatures would be different but still consistent with the scheme of statistical physics.

\section{\label{sec:Con}Conclusions}
There is a lot of discussion about the temperature concept in hadronic physics. One should have in mind, however, that a temperature here is not a self consistent quantity. A little more careful analysis shows that this is just model dependent parameter fitted to experimental data. Within the given class of models, such such thermal models, based on similar assumptions nad the same philosophy one gets similar temperatures when models are applied to explain similar data. There is rather no expectations to build an universal hadronic thermometer which would give the "real" temperature of the hadronic medium. We have a set of different keys fitted to different doors. They are constructed according to the same principles but without hope to create an universal passkey.

\vskip 10mm
\centerline{\bf Acknowledgment}
This work was supported by the Polish National Science Center under
grant no. DEC-2013/10/A/ST2/00106.


\vskip 10mm

\begin{thebibliography}{99}

\bibitem{wigner} \textit{Wigner, E.~P.} \emph{``The Unreasonable Effectiveness of Mathematics in the Natural
Sciences''} Communications in Pure and Applied Mathematics,13, 1-14 (1960)
\bibitem{Koppe:1948az}
  \emph{H.~Koppe}, \emph{"Die Mesonenausbeute beim Beschu{{\ss}} von leichtem Kernen mit $\alpha$-Teilchen"} Zeits.\ f.\ Naturforschung\ \textbf{3(a)}, 251 (1948)

\bibitem{Koppe:1949zz}
  \emph{H.~Koppe}, \emph{"On the Production of Mesons"} Phys.\ Rev.\  {\bf 76}, 688 (1949).

\bibitem{Fermi:1950jd}
  \emph{E.~Fermi},  \emph{"High-energy nuclear events"}, Prog.\ Theor.\ Phys.\  {\bf 5}, 570 (1950)

\bibitem{Heisenberg  : 1949ah}
\emph{W.~Heisenberg}, \emph{"Production of meson showers"} Nature \textbf{164}, 65 (1949)

\bibitem{Hagedorn:1965st}
  R.~Hagedorn,   \emph{``Statistical thermodynamics of strong interactions at high-energies,''}
  Nuovo Cim.\ Suppl.\  {\bf 3}, 147 (1965)

\bibitem{Hagedorn:1970gh}
  R.~Hagedorn, \emph{``Remarks on the thermodynamical model of strong interactions,''}
  Nucl.\ Phys.\ B {\bf 24}, 93 (1970).

\bibitem{brs} For a review see, e.g., P.~Braun-Munzinger, K.~Redlich,
and J.~Stachel, \emph{Quark Gluon Plasma 3}  eds. Hwa~R.~C. and Wang~X.~N. (World
Scientific, Singapore 2004) 491-599;
\end{thebibliography}
\end{document}